\documentclass[reprint,amsmath,amssymb,aps,prl,superscriptaddress,longbibliography]{revtex4-1}
\usepackage{array}
\usepackage{bm}
\usepackage{color}
\usepackage{float}
\usepackage{graphicx}
\usepackage{hyperref}
\hypersetup{colorlinks=true,allcolors=blue}
\usepackage{natbib}
\usepackage{newtxtext}
\usepackage{newtxmath}
\usepackage[caption=false]{subfig}

\begin{document}

\title{Sign structure of thermal Hall conductivity for in-plane-field polarized Kitaev magnets}

\author{Li Ern Chern}
\affiliation{Department of Physics, University of Toronto, Toronto, Ontario M5S 1A7, Canada}

\author{Emily Z. Zhang}
\affiliation{Department of Physics, University of Toronto, Toronto, Ontario M5S 1A7, Canada}

\author{Yong Baek Kim}
\affiliation{Department of Physics, University of Toronto, Toronto, Ontario M5S 1A7, Canada}


\begin{abstract}
The appearance of half-quantized thermal Hall conductivity in $\alpha$-RuCl$_3$ in the presence of in-plane magnetic fields has been taken as a strong evidence for Kitaev spin liquid. Apart from the quantization, the observed sign structure of the thermal Hall conductivity is also consistent with predictions from the exact solution of the Kitaev model. Namely, the thermal Hall conductivity changes sign when the field direction is reversed with respect to the heat current, which is perpendicular to one of the three nearest neighbor bonds on the honeycomb lattice. On the other hand, it is almost zero when the field is applied along the bond direction. Here, we show that such a peculiar sign structure of the thermal Hall conductivity is a generic property of the polarized state in the presence of in-plane magnetic-fields. In this case, thermal Hall effect arises from topological magnons with finite Chern numbers and the sign structure follows from the symmetries of the momentum space Berry curvature. Using a realistic spin model with bond-dependent interactions, we show that the thermal Hall conductivity can have a magnitude comparable to that observed in the experiments. Hence the sign structure alone cannot make a strong case for Kitaev spin liquid. The quantization at very low temperatures, however, will be a decisive test as the magnon contribution vanishes in the zero temperature limit.
\end{abstract}


\maketitle


\textit{Introduction.}---The Kitaev honeycomb model \cite{KITAEV20062}, in which nearest neighbor $S=1/2$ moments are coupled to each other by bond-dependent Ising interactions, is one of the few exactly soluble spin models which lead to an unusual ground state known as quantum spin liquid. In the Kitaev spin liquid, the $S=1/2$ moments fractionalize into Majorana fermions coupled to a $\mathbb{Z}_2$ gauge field. The Kitaev interaction is proposed to have a dominant presence in systems with $4d$/$5d$ transition metal elements \cite{PhysRevLett.102.017205,s42254-019-0038-2} such as Na$_2$IrO$_3$ \cite{PhysRevLett.105.027204,Katukuri_2014,nphys3322} and $\alpha$-RuCl$_3$ \cite{PhysRevB.90.041112,nmat4604,Banerjee1055,s41535-018-0079-2}. However, there exist other interactions as well \cite{PhysRevLett.112.077204}, which pave the way for a zigzag magnetically ordered state \cite{PhysRevB.83.220403,PhysRevB.85.180403,PhysRevB.91.144420,PhysRevB.92.235119} instead of the desired quantum spin liquid. A dramatic twist in the materialization of Kitaev spin liquid came with the observation of half-quantized thermal Hall conductivity in $\alpha$-RuCl$_3$ under an external magnetic field, which has both finite in-plane and out-of-plane components \cite{s41586-018-0274-0}. Since the half-quantization is a signature of Majorana fermions \cite{KITAEV20062,PhysRevLett.119.127204,PhysRevX.8.031032,PhysRevLett.121.147201}, the experiment strongly hints at a field induced Kitaev spin liquid in $\alpha$-RuCl$_3$. If confirmed, this would be a smoking gun that quantum spin liquid does exist in nature, not merely being a theoretical concept.

\begin{figure}
\includegraphics[scale=0.3]{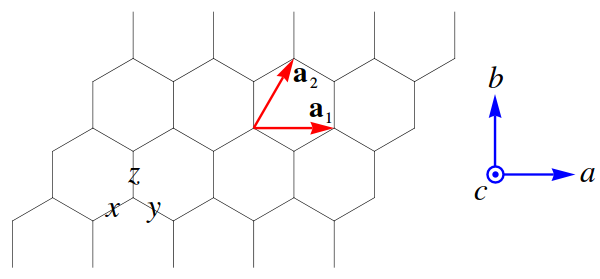}
\caption{\label{convention}The nearest neighbor bond types $x$, $y$, and $z$ in the $K \Gamma \Gamma'$ model, the primitive lattice vectors $\mathbf{a}_1$ and $\mathbf{a}_2$ on the honeycomb lattice, and the crystallographic directions $a$ (in-plane), $b$ (in-plane), and $c$ (out-of-plane). Measured in the cubic basis, according to which the spin components in the $K \Gamma \Gamma'$ model are defined, the $a$, $b$, and $c$ directions are $[11\bar{2}]$, $[\bar{1}10]$, and $[111]$, respectively.}
\end{figure}

More recently, a similar thermal transport measurement with in-plane magnetic fields was performed \cite{2001.01899}. It was reported that the half-quantization of thermal Hall conductivity can still occur even when the field is completely in-plane. Compared to the usual textbook example of two-dimensional conductors where Hall effect only takes place under out-of-plane fields, the sizable thermal Hall signal - not to mention the additional fact that it is half-quantized - in Ref.~\cite{2001.01899} is in some sense anomalous. The experimental setup in Ref.~\cite{2001.01899} is described as follows. The two independent in-plane directions are conventionally chosen to be (i) the $a$ direction which is perpendicular to one of the three nearest neighbor bonds on the honeycomb lattice and (ii) the $b$ direction which is parallel to a nearest neighbor bond and perpendicular to the $a$ direction (see Fig.~\ref{convention}). In the experiment, the heat current is always applied along the $a$ direction, so a finite temperature gradient along the $b$ direction will imply the thermal Hall effect. The magnetic field is applied along the $a$, $b$ and, $-a$ directions, and the observed thermal Hall conductivity is positive, zero, and negative, respectively. Such a sign structure fits into the theory of a non-Abelian spin liquid, which is stabilized in the Kitaev model under a magnetic field. The half-quantized thermal Hall conductivity is observed along the $a$ and $-a$ directions, within a certain range of temperatures and field strengths.

\begin{figure*}
\includegraphics[scale=0.65]{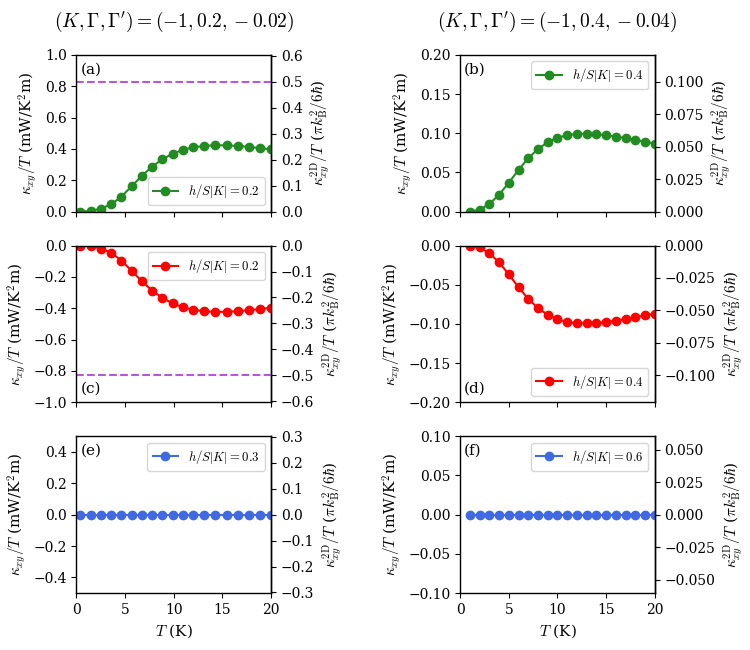}
\caption{\label{thermalhalldata}Thermal Hall conductivity $\kappa_{xy}/T$ due to magnons in various polarized states as a function of temperature $T$, for the parametrizations $(K,\Gamma,\Gamma')=(-1,0.2,-0.02)$ and $(-1,0.4,-0.04)$, shown in the left and right panels respectively. The magnetic field is applied along the $a$ direction in (a) and (b), the $-a$ direction in (c) and (d), and the $b$ direction in (e) and (f). The corresponding values of $\kappa_{xy}^{2\mathrm{D}}/T \equiv \kappa_{xy}d/T$ are also indicated. The purple dashed lines in (a) and (c) represents the half-quantized thermal Hall conductivity.}
\end{figure*}

An independent measurement \cite{ongyoutube} confirmed such a sign structure of the thermal Hall conductivity, but did not quite observe the half-quantization plateau. Rather, the thermal Hall conductivity looks more like a smooth function across a wide range of temperatures (including the suspected spin liquid regime), and vanishes rapidly when the temperature approaches zero. This discovery suggests the existence of a state which may be different from the non-Abelian spin liquid but able to produce the same sign structure of the thermal Hall conductivity.

In this Letter, we theoretically demonstrate that the peculiar sign structure of the thermal Hall conductivity $\kappa_{xy}$ is a generic property of the polarized state in Kitaev magnets under in-plane magnetic fields. In this case, thermal Hall effect arises from topological magnons \cite{1.4959815,PhysRevB.94.094405,Owerre_2017,PhysRevB.98.060404} with finite Chern numbers $C=\pm 1$, while the sign structure of $\kappa_{xy}$ is a consequence of the symmetries of the momentum space Berry curvature. Instead of the Kitaev model, we consider a more realistic $K \Gamma \Gamma'$ model subjected to in-plane magnetic fields as in the experiments \cite{2001.01899,ongyoutube}. We derive analytically the following theorems concerning the sign structure of the thermal Hall conductivity due to magnons in the linear spin wave theory of the polarized state, which are consistent with the experimental observations. \\
\textbf{Theorem 1.}~$\kappa_{xy}$ in the polarized states with the magnetic fields along the $a$ and $-a$ directions differ by a minus sign. \\
\textbf{Theorem 2.}~$\kappa_{xy}$ in the polarized state with the magnetic field along the $b$ direction is zero. \\
Concise proofs will be presented later in the main text, with details relegated to the Supplemental Material \cite{SM}. Theorem 1 only tells us the relative sign of $\kappa_{xy}$ in the $a$ and $-a$ polarized states, not their absolute signs. Therefore, we assume reasonable values of $K$, $\Gamma$, and $\Gamma'$ which minimally models $\alpha$-RuCl$_3$, and perform numerical calculations of $\kappa_{xy}$. We find that, with dominant $K<0$ and $\Gamma>0$, $\kappa_{xy}$ is indeed positive (negative) in the polarized state along $a$ ($-a$) direction, and essentially zero in the polarized state along the $b$ direction, see Figs.~\ref{thermalhalldata}(a)-(f). Moreover, the magnitude and trend of $\kappa_{xy}$ are also comparable to those measured experimentally. Our result suggests that the observed thermal Hall conductivity, in case the half-quantization is absent, may originate from the polarized state with magnons as heat-carriers.

\textit{Model.}---The $K \Gamma \Gamma'$ model, which minimally describes $\alpha$-RuCl$_3$, under a magnetic field is given by $H = \sum_{\lambda \in \lbrace x,y,z \rbrace} \sum_{\langle ij \rangle \in \lambda} \mathbf{S}_i^\mathrm{T} H_\lambda \mathbf{S}_j - \sum_i \mathbf{h} \cdot \mathbf{S}_i$, where
\begin{equation*} \label{hamiltonian}
H_x = \begin{pmatrix} K & \Gamma' & \Gamma' \\ \Gamma' & 0 & \Gamma \\ \Gamma' & \Gamma & 0 \end{pmatrix}, \, H_y = \begin{pmatrix} 0 & \Gamma' & \Gamma \\ \Gamma' & K & \Gamma' \\ \Gamma & \Gamma' & 0 \end{pmatrix}, \, H_z = \begin{pmatrix} 0 & \Gamma & \Gamma' \\ \Gamma & 0 & \Gamma' \\ \Gamma' & \Gamma' & K \end{pmatrix}.
\end{equation*}
We apply the linear spin wave theory \cite{PhysRev.58.1098,Jones_1987} to a field polarized state in the $K \Gamma \Gamma'$ model. First, we rotate the coordinate frames of all spins uniformly such that the $z$-axes align with the spins, $\mathbf{S}_i = R \tilde{\mathbf{S}}_i$. Let the orientation of polarized spins in the original frame be parametrized by two angles $\theta$ and $\phi$ as $\mathbf{S}_i = S (\sin \theta \cos \phi, \sin \theta \sin \phi, \cos \theta)$. We choose the rotation matrix to be \cite{Jones_1987}
\begin{equation} \label{rotationmatrix}
R = \begin{pmatrix} \cos \theta \cos \phi & -\sin \phi & \sin \theta \cos \phi \\ \cos \theta \sin \phi & \cos \phi & \sin \theta \sin \phi \\ -\sin \theta & 0 & \cos \theta \end{pmatrix} \in SO(3) .
\end{equation}
Notice that the columns of $R$ are mutually orthonormal, and they satisfy the right hand rule of cross product. Then, we apply the Holstein Primakoff transformation \cite{PhysRev.58.1098,Jones_1987} to $\tilde{\mathbf{S}}_i$, and neglect the terms of third and higher order in the bosonic operators $b$ and $b^\dagger$. Upon a Fourier transformation, we arrive at the linear spin wave Hamiltonian in momentum space $H/S = \sum_\mathbf{k} \Psi_\mathbf{k}^\dagger \mathrm{D}_\mathbf{k} \Psi_\mathbf{k}$, where $\mathrm{D}_\mathbf{k}$ is a four-dimensional Hermitian matrix and $\Psi_\mathbf{k} = (b_{1\mathbf{k}}, b_{2\mathbf{k}}, b_{1-\mathbf{k}}^\dagger, b_{2-\mathbf{k}}^\dagger)$. $\mathrm{D}_\mathbf{k}$ has to be diagonalized by a Bogoliubov transformation in order to preserve the commutation relation of bosons. Once we obtain the linear spin wave dispersion $\varepsilon_{n \mathbf{k}}$, we can calculate the thermal Hall conductivity due to magnons \cite{PhysRevLett.106.197202,PhysRevB.89.054420,JPSJ.86.011010}
\begin{equation} \label{thermalhallformula}
\kappa_{xy} = - \frac{k_\mathrm{B}^2 T}{\hbar V} \sum_n \sum_{\mathbf{k} \in \mathrm{FBZ}} \left \lbrace c_2 \left[ g \left( \varepsilon_{n\mathbf{k}} \right) \right] - \frac{\pi^2}{3} \right \rbrace \Omega_{n \mathbf{k}},
\end{equation}
where FBZ denotes the first Brillouin zone, $c_2 (x) = (1+x) \lbrace \ln [(1+x) / x] \rbrace^2 - (\ln x)^2 - 2 \mathrm{Li}_2 (-x)$, $g$ is the Bose-Einstein distribution, and $\Omega_{n \mathbf{k}}$ is the Berry curvature of the $n$th band at momentum $\mathbf{k}$. The term $-\pi^2/3$ will be dropped in subsequent discussions because the summation of all Chern numbers is zero \cite{PhysRevB.87.174427}.

\textit{Proof of Theorem 1.}---The sets of angles $\lbrace \theta , \phi \rbrace$ in \eqref{rotationmatrix} for the polarized states along the $a$ and $-a$ directions are, respectively, $\lbrace \cos^{-1} (-\sqrt{2/3}) , \pi / 4 \rbrace$ and $\lbrace \cos^{-1} (\sqrt{2/3}) , 5 \pi / 4 \rbrace$. After some algebra \cite{SM}, one can show that the linear spin wave Hamiltonians of the $a$ and $-a$ polarized states are related by $\mathrm{D}_\mathbf{k}^{\bar{a}}=(\mathrm{D}_{-\mathbf{k}}^a)^*$. Suppose that $\mathrm{T}_\mathbf{k}^a$ is the Bogoliubov transformation of $\mathrm{D}_\mathbf{k}^a$, then $\mathrm{T}_\mathbf{k}^{\bar{a}} = (\mathrm{T}_{-\mathbf{k}}^a)^*$ is the Bogoliubov transformation of $\mathrm{D}_\mathbf{k}^{\bar{a}}$,
\begin{subequations}
\begin{align}
& (\mathrm{T}_\mathbf{k}^{\bar{a}})^\dagger \mathrm{D}_\mathbf{k}^{\bar{a}} \mathrm{T}_\mathbf{k}^{\bar{a}} = \left[ \left( \mathrm{T}_{-\mathbf{k}}^a \right)^\dagger \mathrm{D}_{-\mathbf{k}}^a \mathrm{T}_{-\mathbf{k}}^a \right]^* = \mathcal{E}_{-\mathbf{k}}^a =  \mathcal{E}^{\bar{a}}_\mathbf{k}, \label{energymatrixminusa} \\
& \mathrm{T}_\mathbf{k}^{\bar{a}} \sigma^3 \left( \mathrm{T}_\mathbf{k}^{\bar{a}} \right)^\dagger = \left[ \mathrm{T}_{-\mathbf{k}}^a \sigma^3 \left( \mathrm{T}_{-\mathbf{k}}^a \right)^\dagger \right]^* = \sigma^3 ,
\end{align}
\end{subequations}
where $\sigma^3 = \mathrm{diag} \left( 1,1,-1,-1 \right)$. \eqref{energymatrixminusa} says that the energy eigenvalues at $\mathbf{k}$ when $\mathbf{h} \parallel -a$ are the same as those at $-\mathbf{k}$ when $\mathbf{h} \parallel a$. Next, we relate the Berry curvatures in the two polarized states,
\begin{equation}
\begin{aligned}[b]
\Omega_{n -\mathbf{k}}^{\bar{a}} &= i \left[ \sigma^3 \left( \frac{\partial \mathrm{T}_{-\mathbf{k}}^{\bar{a}}}{\partial k_x} \right)^\dagger \sigma^3 \frac{\partial \mathrm{T}_{-\mathbf{k}}^{\bar{a}}}{\partial k_y} - \left( x \longleftrightarrow y \right) \right]_{nn} \\
&= - \left \lbrace i \left[ \sigma^3 \left( \frac{\partial \mathrm{T}_{\mathbf{k}}^a}{\partial k_x} \right)^\dagger \sigma^3 \frac{\partial \mathrm{T}_{\mathbf{k}}^a}{\partial k_y} - \left( x \longleftrightarrow y \right) \right]_{nn} \right \rbrace^* \\
&= - \Omega_{n \mathbf{k}}^a .
\end{aligned}
\end{equation}
Therefore, the thermal Hall conductivities are related by
\begin{equation}
\begin{aligned}[b]
\kappa_{xy}^{\bar{a}} &= - \frac{k_\mathrm{B}^2 T}{\hbar V} \sum_{n \mathbf{k}} c_2 \left[ g \left( \varepsilon_{n -\mathbf{k}}^{\bar{a}} \right) \right] \Omega_{n -\mathbf{k}}^{\bar{a}} \\
&= \frac{k_\mathrm{B}^2 T}{\hbar V} \sum_{n \mathbf{k}} c_2 \left[ g \left( \varepsilon_{n \mathbf{k}}^a \right) \right] \Omega_{n \mathbf{k}}^a \\
&= - \kappa_{xy}^a .
\end{aligned}
\end{equation}
This completes the proof of Theorem 1.

\textit{Proof of Theorem 2.}---The set of angles $\lbrace \theta , \phi \rbrace$ in \eqref{rotationmatrix} for the polarized state along the $b$ direction is $\lbrace \pi / 2 , 3 \pi / 4 \rbrace$. After some algebra \cite{SM}, one can show that the linear spin wave Hamiltonian of the $b$ polarized state $\mathrm{D}_\mathbf{k}^b$ depends on $k_x$ and $k_y$ only through the combination $2 \cos (k_x a/2) \exp ( -i \sqrt{3} k_y a / 2 )$ and its complex conjugate. Therefore, the linear spin wave Hamiltonian is an even function in $k_x$, $\mathrm{D}_{(-k_x,k_y)}^b = \mathrm{D}_{(k_x, k_y)}^b$. Consequently, the matrix of energy eigenvalues and the Bogoliubov transformation are even in $k_x$, $\mathcal{E}_{(-k_x,k_y)}^b = \mathcal{E}_{(k_x, k_y)}^b$ and $\mathrm{T}_{(-k_x,k_y)}^b = \mathrm{T}_{(k_x, k_y)}^b$. Since the derivative of an even function is an odd function,
\begin{equation} \label{derivativeoddeven}
\frac{\partial \mathrm{T}_{(-k_x,k_y)}^b}{\partial k_x} = - \frac{\partial \mathrm{T}_{(k_x,k_y)}^b}{\partial k_x} .
\end{equation}
In addition, we have
\begin{equation}
\begin{aligned}[b]
\frac{\partial \mathrm{T}_{(-k_x,k_y)}^b}{\partial k_y} &= \lim_{\epsilon \longrightarrow 0} \frac{\mathrm{T}_{(-k_x,k_y + \epsilon)}^b - \mathrm{T}_{(-k_x,k_y)}^b}{\epsilon} \\
&= \lim_{\epsilon \longrightarrow 0} \frac{\mathrm{T}_{(k_x,k_y + \epsilon)}^b - \mathrm{T}_{(k_x,k_y)}^b}{\epsilon} \\
&= \frac{\partial \mathrm{T}_{(k_x,k_y)}^b}{\partial k_y} .
\end{aligned}
\end{equation}
Therefore, the Berry curvatures at $\mathbf{k}=(\pm k_x,k_y)$ are related by
\begin{equation} \label{curvaturerelationb}
\begin{aligned}[b]
\Omega_{n (-k_x,k_y)}^b &= i \left[ \sigma^3 \left( \frac{\partial \mathrm{T}^b_{(-k_x,k_y)}}{\partial k_x} \right)^\dagger \sigma^3 \frac{\partial \mathrm{T}_{(-k_x,k_y)}^b}{\partial k_y} - \left( x \longleftrightarrow y \right) \right]_{nn} \\
&= - i \left[ \sigma^3 \left( \frac{\partial \mathrm{T}^b_{(k_x,k_y)}}{\partial k_x} \right)^\dagger \sigma^3 \frac{\partial \mathrm{T}_{(k_x,k_y)}^b}{\partial k_y} - \left( x \longleftrightarrow y \right) \right]_{nn} \\
&= - \Omega_{n (k_x,k_y)}^b ,
\end{aligned}
\end{equation}
which leads to zero thermal Hall conductivity, because the first Brillouin zone (a hexagon centered at $\mathbf{k}=0$ in the reciprocal space) is symmetric about $k_x = 0$. This completes the proof of Theorem 2. \eqref{curvaturerelationb} also implies that the Chern numbers $C_n \sim \sum_\mathbf{k} \Omega_{n \mathbf{k}}$ of the magnon bands in the $b$ polarized state are zero.

The fact that $\mathrm{D}_\mathbf{k}^b$ is even in $k_x$ can be argued more heuristically without scrutinizing its explicit form, as follows. When the field is applied along the $b$ direction, the spin Hamiltonian possesses a $C_2$ rotational symmetry about the $b$ axis \cite{PhysRevResearch.2.013072,2004.13723}. In principle, $C_2$ acts on both the spatial coordinates and the spins, $C_2: \mathbf{S}_i \longrightarrow C_2^{-1}\mathbf{S}_{C_2(i)}$. However, in the $b$ polarized state, the spin rotation part is effectively an identity operator since all spins lie exactly along the axis of rotation. In other words, $C_2$ only affects the spatial coordinates. In the reciprocal space, the $C_2$ symmetry translates into the invariance of $\mathrm{D}_\mathbf{k}^b$ under $k_x \longrightarrow -k_x$, and the rest of the proof follows.

\textit{Corollary.}---Theorems 1 and 2 still hold if Heisenberg interactions, for instance $J$ ($J_3$) between the (third) nearest neighbors, are added to the $K \Gamma \Gamma'$ model, due to the following reasons. (i) The Heisenberg interaction is proportional to an identity matrix, which is left invariant under a global rotation of spins. (ii) The Heisenberg interaction is isotropic, namely it is the same along all bond directions. This indicates the robustness of Theorems 1 and 2 against the choice of spin model for $\alpha$-RuCl$_3$, which is suggested to be a $J K \Gamma \Gamma'$ model \cite{PhysRevB.93.155143} or a $J K \Gamma J_3$ model \cite{PhysRevB.93.214431,PhysRevB.96.115103,s41467-017-01177-0}.

\begin{figure}
\begin{centering}
\includegraphics[scale=0.54]{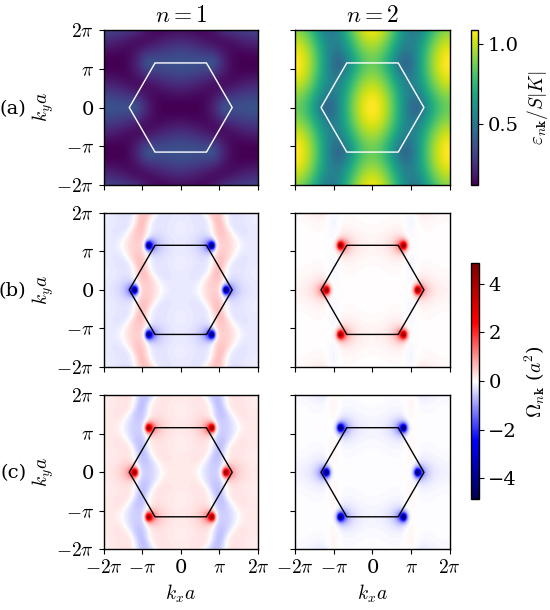}
\par\end{centering}
\caption{\label{berry_a}(a) Linear spin wave dispersion $\varepsilon_{n \mathbf{k}}$ of the $a$ and $-a$ polarized states, and Berry curvature $\Omega_{n \mathbf{k}}$ in (b) the $a$ polarized state and (c) the $-a$ polarized state, as functions of momentum $\mathbf{k}$. $n=1$ and $2$ are the band indices. The interaction parameters are chosen to be $K=-1$, $\Gamma=0.2$, and $\Gamma'=-0.02$. The first Brillouin zone is indicated by a hexagon. The symbol $a$ that appears in the units of $\mathbf{k}$ and $\Omega_{n \mathbf{k}}$ is the lattice constant, not to be confused with the $a$ direction.}
\end{figure}

\textit{Numerical Results.}---Using two sets of interaction parameters $(K,\Gamma,\Gamma')=(-1,0.2,-0.02)$ and $(-1,0.4,-0.04)$ that are relevant to $\alpha$-RuCl$_3$, we numerically evaluate the thermal Hall conductivity due to magnons \eqref{thermalhallformula} in the $a$, $b$, and $-a$ polarized states. We first use classical simulated annealing \cite{PhysRevLett.117.277202,PhysRevResearch.2.013014} to obtain the critical fields to the polarized states along the $a$, $b$, and $-a$ directions. We set the spin magnitude to be $S=1/2$ in the linear spin wave theory, and assume the strength of Kitaev interaction to be $\lvert K \rvert = 80 \, \mathrm{K}$ \cite{PhysRevB.93.155143}, in the calculation of the thermal Hall conductivity. We also use the interlayer distance $d=5.72 \, \textrm{\AA}$ of $\alpha$-RuCl$_3$ \cite{s41586-018-0274-0}.

\begin{figure}
\begin{centering}
\includegraphics[scale=0.54]{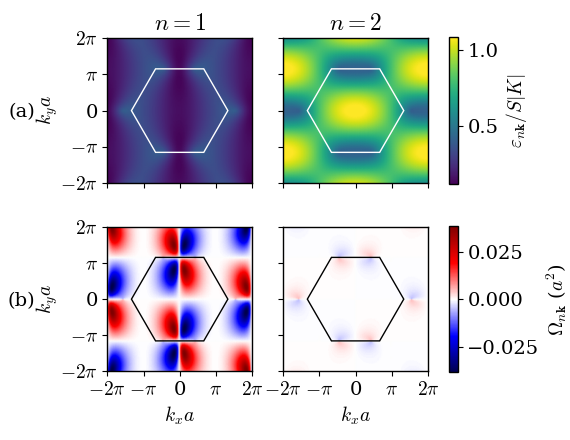}
\par\end{centering}
\caption{\label{berry_b}(a) Linear spin wave dispersion $\varepsilon_{n \mathbf{k}}$ and (b) Berry curvature $\Omega_{n \mathbf{k}}$ of the $b$ polarized state as functions of momentum $\mathbf{k}$. The interaction parameters and notations used are same as in Figs.~\ref{berry_a}(a)-(c).}
\end{figure}

We plot the thermal Hall conductivity as a function of temperature for the three polarized states in Figs.~\ref{thermalhalldata}(a)-(f). The field strengths $h$ are chosen such that the system is indeed in the corresponding polarized states according to the classical model. We make two important observations from the results. First, $\kappa_{xy}$ is positive, zero, and negative when the field is along the $a$, $b$, and $-a$ directions, respectively, which matches the experimentally observed signs of the thermal Hall conductivity. The sign structure is also consistent with Theorems 1 and 2. Second, in the $a$ and $-a$ polarized states, $\kappa_{xy}/T$ is of the order $0.1 \times 10^{-3} \, \mathrm{W} / \mathrm{K}^2 \mathrm{m}$, which is comparable in magnitude to the thermal Hall signals experimentally measured at low temperatures \cite{s41586-018-0274-0,2001.01899,ongyoutube}. For instance, the maximum value of $\kappa_{xy}^{2\mathrm{D}}/T$ for $(K,\Gamma,\Gamma')=(-1,0.2,-0.02)$ is about $0.25 \times (\pi/6) (k_\mathrm{B}^2/\hbar)$, which is half of the half-quantized value.

Furthermore, we numerically verify the symmetries discussed in the proofs of Theorems 1 and 2, by plotting the linear spin wave dispersion and the Berry curvature as functions of momentum \cite{PhysRevB.98.060412,PhysRevResearch.1.013014} in the polarized states. When the field is flipped from the $a$ direction to the $-a$ direction, the dispersion remains the same, as shown in Fig.~\ref{berry_a}(a). However, the Berry curvature changes sign, as shown in Figs.~\ref{berry_a}(b) and (c). By \eqref{thermalhallformula}, the thermal Hall conductivity gains an overall minus sign. When the field is along the $\pm a$ direction, the Chern numbers of the lower ($n=1$) and upper ($n=2$) magnon bands are $\mp 1$ and $\pm 1$ respectively, signifying their topological nontriviality. On the other hand, when the field is applied along the $b$ direction, the linear dispersion is symmetric about $k_x=0$, as shown in Fig.~\ref{berry_b}(a). However, the Berry curvature is antisymmetric about $k_x=0$, as shown in Figs.~\ref{berry_b}(b), which results in cancellations of the summands in \eqref{thermalhallformula}, eventually leading to a zero thermal Hall conductivity. The magnon bands are topologically trivial, i.e.~they have zero Chern numbers.

\textit{Conclusion.}---In this Letter, we have shown both analytically and numerically that the thermal Hall conductivity due to magnons in the field polarized states along the $a$, $b$, and $-a$ directions is positive, zero, and negative, respectively, which agrees with the experimentally observed sign structure \cite{2001.01899,ongyoutube} in the Kitaev material $\alpha$-RuCl$_3$. If the half-quantization plateau is present (absent), then the ground state may be the non-Abelian spin liquid (polarized state) with Majorana fermions (magnons). Most importantly, thermal Hall effect does occur in both the non-Abelian spin liquid and the polarized state, and their thermal Hall conductivities have the same sign structure. Therefore, the sign structure alone cannot serve as a strong evidence for the non-Abelian spin liquid. The ultimate test for the non-Abelian spin liquid will be the half-quantization of thermal Hall conductivity at very low temperatures, where the magnon contribution vanishes.

\begin{acknowledgments}
We thank Hae-Young Kee for useful discussions. L.E.C.~was supported by the Ontario Graduate Scholarship. E.Z.Z.~and Y.B.K. were supported by the NSERC of Canada. Y.B.K.~was further supported by the Killam Research Fellowship from the Canada Council for the Arts and the Center for Quantum Materials at the University of Toronto. Part of the computations were performed on the Cedar and Niagara \cite{1742-6596-256-1-012026} clusters, which are hosted by WestGrid and SciNet in partnership with Compute Canada.
\end{acknowledgments}

\bibliography{reference200827}

\clearpage

\onecolumngrid

\begin{center}
\textbf{\large Supplementary Materials: \\ Sign structure of thermal Hall conductivity for in-plane-field polarized Kitaev magnets}
\end{center}
\begin{center}
Li Ern Chern$^1$, Emily Z. Zhang$^1$, and Yong Baek Kim$^1$
\end{center}
\begin{center}
{\small
\textit{$^1$Department of Physics, University of Toronto, Toronto, Ontario M5S 1A7, Canada}}
\end{center}

\setcounter{equation}{0}
\setcounter{figure}{0}
\setcounter{table}{0}
\setcounter{page}{1}

\renewcommand{\thesection}{S\arabic{section}}
\renewcommand{\theequation}{S\arabic{equation}}
\renewcommand{\thefigure}{S\arabic{figure}}
\renewcommand{\thetable}{S\arabic{table}}

\section{Details of Analytical Proofs}

Here we provide more details to the analytical proofs of the two theorems concerning the sign structure of the magnon thermal Hall conductivity $\kappa_{xy}$ in the $K \Gamma \Gamma'$ model, which are the main results of this study. \\
\textbf{Theorem 1.} $\kappa_{xy}$ in the polarized states with the magnetic fields along the $a$ and $-a$ directions differ by a minus sign. \\
\textbf{Theorem 2.} $\kappa_{xy}$ in the polarized state with the magnetic field along the $b$ direction is zero. \\

\noindent \textbf{Preliminaries.} The $K \Gamma \Gamma'$ model under a magnetic field is given in the main text. To derive the linear spin wave Hamiltonian of a polarized state, the coordinate frames of all spins are first rotated uniformly such that the $z$-axes align with the spins, $\mathbf{S}_i = R \tilde{\mathbf{S}}_i$. In the rotated frame, the Hamiltonian is $H =\sum_{\lambda \in \lbrace x,y,z \rbrace} \sum_{\langle ij \rangle \in \lambda} \tilde{\mathbf{S}}_i^\mathrm{T} \tilde{H}_\lambda \tilde{\mathbf{S}}_j - \sum_i \tilde{\mathbf{h}} \cdot \tilde{\mathbf{S}}_i$, where $\tilde{H}_\lambda = R^\mathrm{T} H_\lambda R$ and $\tilde{\mathbf{h}}=R^\mathrm{T} \mathbf{h}$. We choose the rotation matrix $R \in SO(3)$ according to (1). For convenience of latter discussion, we introduce the notation
\begin{equation}
\tilde{H}_\lambda = \begin{pmatrix} \tilde{h}_{11}^\lambda & \tilde{h}_{12}^\lambda & \tilde{h}_{13}^\lambda \\ \tilde{h}_{21}^\lambda & \tilde{h}_{22}^\lambda & \tilde{h}_{23}^\lambda \\ \tilde{h}_{31}^\lambda & \tilde{h}_{32}^\lambda & \tilde{h}_{33}^\lambda \end{pmatrix}
\end{equation}
and notice that $\tilde{H}_\lambda^\mathrm{T} = \tilde{H}_\lambda$. Then, we apply the Holstein Primakoff transformation,
\begin{subequations}
\begin{align}
\tilde{S}_i^z &= S - b_i^\dagger b_i = S - n_i , \\
\tilde{S}_i^x &= \frac{\sqrt{2 S - n_i}b_i + b_i^\dagger \sqrt{2 S - n_i}}{2} , \\
\tilde{S}_i^y &= \frac{\sqrt{2 S - n_i}b_i - b_i^\dagger \sqrt{2 S - n_i}}{2i} ,
\end{align}
\end{subequations}
and keep only terms up to second order in the bosonic operators $b$ and $b^\dagger$. A subsequent Fourier transformation yields the linear spin wave Hamiltonian in momentum space, $H/S = \sum_\mathbf{k} \Psi_\mathbf{k}^\dagger \mathrm{D}_\mathbf{k} \Psi_\mathbf{k}$, where $\mathrm{D}_\mathbf{k}$ is a four dimensional Hermitian matrix and $\Psi_\mathbf{k} = (b_{1\mathbf{k}},b_{2\mathbf{k}},b_{1-\mathbf{k}}^\dagger,b_{2-\mathbf{k}}^\dagger)$. $\mathrm{D}_\mathbf{k}$ assumes the form
\begin{subequations}
\begin{align}
\mathrm{D}_\mathbf{k} &= \begin{pmatrix} \mathrm{A}_\mathbf{k} & \mathrm{B}_\mathbf{k} \\ \mathrm{B}_{-\mathbf{k}}^* & \mathrm{A}_{-\mathbf{k}}^\mathrm{T} \end{pmatrix} , \\
\mathrm{A}_\mathbf{k} &= \begin{pmatrix} h & 0 \\ 0 & h \end{pmatrix} + \sum_\lambda \begin{pmatrix} - \tilde{h}_{33}^\lambda & \frac{1}{2} \left(\tilde{h}_{11}^\lambda + \tilde{h}_{22}^\lambda \right) e^{i \mathbf{k} \cdot \delta_\lambda} \\  \frac{1}{2} \left(\tilde{h}_{11}^\lambda + \tilde{h}_{22}^\lambda \right) e^{-i \mathbf{k} \cdot \delta_\lambda} & - \tilde{h}_{33}^\lambda \end{pmatrix} , \\
\mathrm{B}_\mathbf{k} &= \sum_\lambda \frac{1}{2} \left( \tilde{h}_{11}^\lambda + 2 i \tilde{h}_{12}^\lambda - \tilde{h}_{22}^\lambda \right) \begin{pmatrix} 0 & e^{i \mathbf{k} \cdot \delta_\lambda} \\ e^{-i \mathbf{k} \cdot \delta_\lambda} & 0 \end{pmatrix} .
\end{align}
\end{subequations}

Some remarks are in order. First, the rotated hamiltonian (S1) is real. Second, only the entries $\tilde{h}_{11}^\lambda$, $\tilde{h}_{12}^\lambda = \tilde{h}_{21}^\lambda$, $\tilde{h}_{22}^\lambda$, and $\tilde{h}_{33}^\lambda$ in (S1) are relevant to the linear spin wave Hamiltonian, as shown in (S3a)-(S3c). Third, only $\tilde{h}_{12}^\lambda$ and $\tilde{h}_{21}^\lambda$ are multiplied by the imaginary unit $i$ in $\mathrm{D}_\mathbf{k}$, as shown in (S3a)-(S3c).

Finally, the reciprocal lattice vectors $\mathbf{b}_1 = (4 \pi / \sqrt{3} a) [\cos (\pi/6) \hat{\mathbf{x}} - \sin (\pi/6) \hat{\mathbf{y}}]$ and $\mathbf{b}_2 = (4 \pi / \sqrt{3} a) \hat{\mathbf{y}}$ are defined through the primitive lattice vectors $\mathbf{a}_1 = a \hat{\mathbf{x}}$ and $\mathbf{a}_2=a [\cos (\pi/3) \hat{\mathbf{x}} + \sin (\pi/3) \hat{\mathbf{y}}]$, where $\hat{\mathbf{x}}$ and $\hat{\mathbf{y}}$ are unit vectors in the $a$ and $b$ directions respectively. The first Brillouin zone is equivalent to the following parallelogram in reciprocal space,
\begin{equation}
\mathbf{k} = k_1 \hat{\mathbf{b}}_1 + k_2 \hat{\mathbf{b}}_2, \, k_1 \in [0, 2 \pi), \, k_2 \in [0, 2 \pi) ; \, \hat{\mathbf{b}}_1 \equiv \frac{2}{\sqrt{3}a} \left( \cos \frac{\pi}{6} \hat{\mathbf{x}} - \sin \frac{\pi}{6} \hat{\mathbf{y}} \right) , \,
\hat{\mathbf{b}}_2 \equiv \frac{2}{\sqrt{3}a} \hat{\mathbf{y}} .
\end{equation}
If we write $\mathbf{k}=k_x \hat{\mathbf{x}} + k_y \hat{\mathbf{y}}$, then $(k_1,k_2)$ and $(k_x,k_y)$ are related by a linear transformation,
\begin{equation}
\begin{pmatrix} k_1 \\ k_2 \end{pmatrix} = a \begin{pmatrix} 1 & 0 \\ \sin \frac{\pi}{6} & \cos \frac{\pi}{6} \end{pmatrix} \begin{pmatrix} k_x \\ k_y \end{pmatrix} .
\end{equation}

\noindent \textbf{Proof of Theorem 1.} The rotation matrices (1) for the polarized states along the $a$ and $-a$ directions are, respectively,
\begin{subequations}
\begin{align}
R^a &= \begin{pmatrix} - \frac{1}{\sqrt{3}} & - \frac{1}{\sqrt{2}} & \frac{1}{\sqrt{6}} \\ - \frac{1}{\sqrt{3}} & \frac{1}{\sqrt{2}} & \frac{1}{\sqrt{6}} \\ - \frac{1}{\sqrt{3}} & 0 & - \frac{2}{\sqrt{6}} \end{pmatrix} , \\
R^{\bar{a}} &= \begin{pmatrix} - \frac{1}{\sqrt{3}} & \frac{1}{\sqrt{2}} & - \frac{1}{\sqrt{6}} \\ - \frac{1}{\sqrt{3}} & - \frac{1}{\sqrt{2}} & - \frac{1}{\sqrt{6}} \\ - \frac{1}{\sqrt{3}} & 0 & \frac{2}{\sqrt{6}} \end{pmatrix} .
\end{align}
\end{subequations}
The explicit forms of the rotated spin Hamiltonians (S1) of the $a$ and $-a$ polarized states are, repectively,
\begin{subequations}
\begin{align}
& \tilde{H}_x^a = \begin{pmatrix} \frac{K + 2 \Gamma + 4 \Gamma'}{3} & \frac{K - \Gamma + \Gamma'}{\sqrt{6}} & \frac{-K + \Gamma - \Gamma'}{3 \sqrt{2}} \\ \frac{K - \Gamma + \Gamma'}{\sqrt{6}} & \frac{K - 2 \Gamma'}{2} & \frac{-K - 2\Gamma + 2 \Gamma'}{2 \sqrt{3}} \\ \frac{-K + \Gamma - \Gamma'}{3 \sqrt{2}} & \frac{-K - 2\Gamma + 2 \Gamma'}{2 \sqrt{3}} & \frac{K - 4\Gamma - 2 \Gamma'}{6} \end{pmatrix} , \, \tilde{H}_y^a = \begin{pmatrix} \frac{K + 2 \Gamma + 4 \Gamma'}{3} & \frac{-K + \Gamma - \Gamma'}{\sqrt{6}} & \frac{-K + \Gamma - \Gamma'}{3 \sqrt{2}} \\ \frac{-K + \Gamma - \Gamma'}{\sqrt{6}} & \frac{K - 2 \Gamma'}{2} & \frac{K + 2\Gamma - 2 \Gamma'}{2 \sqrt{3}} \\ \frac{-K + \Gamma - \Gamma'}{3 \sqrt{2}} & \frac{K + 2 \Gamma - 2 \Gamma'}{2 \sqrt{3}} & \frac{K - 4 \Gamma - 2 \Gamma'}{6} \end{pmatrix} , \, \tilde{H}_z^a = \begin{pmatrix} \frac{K + 2 \Gamma + 4 \Gamma'}{3} & 0 & \frac{\sqrt{2}(K - \Gamma + \Gamma')}{3} \\ 0 & -\Gamma & 0 \\ \frac{\sqrt{2}(K - \Gamma + \Gamma')}{3} & 0 & \frac{2 K + \Gamma - 4 \Gamma'}{3} \end{pmatrix} , \\
& \tilde{H}_x^{\bar{a}} = \begin{pmatrix} \frac{K + 2 \Gamma + 4 \Gamma'}{3} & \frac{-K + \Gamma - \Gamma'}{\sqrt{6}} & \frac{K - \Gamma + \Gamma'}{3 \sqrt{2}} \\ \frac{-K + \Gamma - \Gamma'}{\sqrt{6}} & \frac{K - 2 \Gamma'}{2} & \frac{-K - 2\Gamma + 2 \Gamma'}{2 \sqrt{3}} \\ \frac{K - \Gamma + \Gamma'}{3 \sqrt{2}} & \frac{-K - 2\Gamma + 2 \Gamma'}{2 \sqrt{3}} & \frac{K - 4 \Gamma - 2 \Gamma'}{6} \end{pmatrix} , \, \tilde{H}_y^{\bar{a}} = \begin{pmatrix} \frac{K + 2 \Gamma + 4 \Gamma'}{3} & \frac{K - \Gamma + \Gamma'}{\sqrt{6}} & \frac{K - \Gamma + \Gamma'}{3 \sqrt{2}} \\ \frac{K - \Gamma + \Gamma'}{\sqrt{6}} & \frac{K - 2 \Gamma'}{2} & \frac{K + 2 \Gamma - 2 \Gamma'}{2 \sqrt{3}} \\ \frac{K - \Gamma + \Gamma'}{3 \sqrt{2}} & \frac{K + 2\Gamma - 2 \Gamma'}{2 \sqrt{3}} & \frac{K - 4 \Gamma - 2 \Gamma'}{6} \end{pmatrix} , \, \tilde{H}_z^{\bar{a}} = \begin{pmatrix} \frac{K + 2 \Gamma + 4 \Gamma'}{3} & 0 & \frac{\sqrt{2}(-K + \Gamma - \Gamma')}{3} \\ 0 & -\Gamma & 0 \\ \frac{\sqrt{2}(-K + \Gamma - \Gamma')}{3} & 0 & \frac{2 K + \Gamma - 4 \Gamma'}{3} \end{pmatrix} ,
\end{align}
\end{subequations}
while the Zeeman term has the same form for any polarized state, regardless of the field direction. The most important feature in (S7a) and (S7b) is that, among the entries relevant to the linear spin wave Hamiltonian, $\tilde{h}_{12}^\lambda$ and $\tilde{h}_{21}^\lambda$ change sign, while $\tilde{h}_{11}^\lambda$, $\tilde{h}_{22}^\lambda$, and $\tilde{h}_{33}^\lambda$ remain invariant, when the field direction is flipped from $a$ to $-a$. (S3a)-(S3c) then imply that the linear spin wave Hamiltonians of the $a$ and $-a$ polarized states are related by $\mathrm{D}_\mathbf{k}^{\bar{a}} = (\mathrm{D}_{-\mathbf{k}}^a)^*$. The rest of the proof is contained in the main text. \\

\noindent \textbf{Proof of Theorem 2.} The rotation matrix (1) for the polarized state along the $b$ direction is
\begin{equation}
R^b = \begin{pmatrix} 0 & - \frac{1}{\sqrt{2}} & - \frac{1}{\sqrt{2}} \\ 0 & - \frac{1}{\sqrt{2}} & \frac{1}{\sqrt{2}} \\ -1 & 0 & 0 \end{pmatrix}
\end{equation}
The explicit form of the rotated spin Hamiltonian (S1) of the $b$ polarized state is
\begin{equation}
\tilde{H}_x^b = \begin{pmatrix} 0 & \frac{\Gamma + \Gamma'}{\sqrt{2}} & \frac{-\Gamma + \Gamma'}{\sqrt{2}} \\ \frac{\Gamma + \Gamma'}{\sqrt{2}} & \frac{K + 2 \Gamma'}{2} & \frac{K}{2} \\ \frac{-\Gamma + \Gamma'}{\sqrt{2}} & \frac{K}{2} & \frac{K - 2 \Gamma'}{2} \end{pmatrix}, \, \tilde{H}_y^b = \begin{pmatrix} 0 & \frac{\Gamma + \Gamma'}{\sqrt{2}} & \frac{\Gamma - \Gamma'}{\sqrt{2}} \\ \frac{\Gamma + \Gamma'}{\sqrt{2}} & \frac{K + 2 \Gamma'}{2} & -\frac{K}{2} \\ \frac{\Gamma - \Gamma'}{\sqrt{2}} & -\frac{K}{2} & \frac{K - 2 \Gamma'}{2} \end{pmatrix}, \, \tilde{H}^b_z = \begin{pmatrix} K & \sqrt{2} \Gamma' & 0 \\ \sqrt{2} \Gamma' & \Gamma & 0 \\ 0 & 0 & - \Gamma \end{pmatrix}.
\end{equation}
The most important feature in (S9) is that, among the matrix elements relevant to the linear spin wave Hamitlonian, $\tilde{h}_{ij}^x = \tilde{h}_{ij}^y$. On the other hand, the displacements (between unit cells) along the $x$, $y$, and $z$ bonds are $\delta_x = (0,-1)$, $\delta_y = (1,-1)$, and $\delta_z=(0,0)$, as measured by the primitive lattice vectors $\mathbf{a}_1$ and $\mathbf{a}_2$ (see Fig.~1). Therefore, the phase factors in (S3b) and (S3c) are momentum dependent only along the $x$ and $y$ bonds. They are $e^{-i k_2}$ and $e^{i(k_1-k_2)}$ (and their complex conjugates). By (S9), these two phase factors share the same coefficient in the linear spin wave Hamiltonian (S3a) of the $b$ polarized state. We may as well say that $D_\mathbf{k}^b$ depends on $k_1$ and $k_2$ only through the combination $e^{-i k_2} + e^{i (k_1-k_2)} = 2 \cos (k_x a / 2) e^{-i \sqrt{3} k_y a / 2}$ (and its complex conjugate). The rest of the proof is contained in the main text.
\end{document}